\documentclass[11pt,a4paper]{article}

\usepackage[utf8]{inputenc}
\usepackage[T1]{fontenc}
\usepackage[english]{babel}
\usepackage{geometry}
\usepackage{authblk}
\usepackage{amsmath,amssymb}
\usepackage{hyperref}
\usepackage{graphicx}
\usepackage{subcaption}
\usepackage{tikz}
\usepackage{float}
\usepackage[numbers,sort&compress]{natbib}

\title{\textbf{A Mid-Infrared Absorber Inspired by the Wing-Scale Photonic Structure of \textit{Troides magellanus}}}

\author[1]{Amandine Marchand}
\author[2]{Kevin Delmote}
\author[2]{Olivier Deparis}
\author[1,2,3,4]{Sébastien R. Mouchet}

\affil[1]{\textit{Micro- and Nanophotonic Materials Group and Research Institute for Materials Science and
Engineering, University of Mons, Place du Parc 20, 7000 Mons, Belgium}}
\affil[2]{\textit{Department of Physics and Namur Institute of Structured Matter (NISM), University of
Namur, Rue de Bruxelles 61, 5000 Namur, Belgium}}
\affil[3]{\textit{Institute of Life, Earth and Environment (ILEE), University of Namur, Rue de Bruxelles 61,
5000 Namur, Belgium}}
\affil[4]{\textit{Department of Physics and Astronomy, University of Exeter, Stocker Road, Exeter EX4 4QL,
United Kingdom}}
\date{}

\begin{document}

\maketitle

\begin{abstract}
Growing environmental and energy-related challenges have intensified the search for technologies capable of improving energy recovery and resource efficiency. Bioinspired photonic structures offer promising opportunities for the development of efficient infrared absorbers by exploiting optical mechanisms found in natural systems. In this work, we present a bioinspired approach for the design and optimization of a broadband mid-infrared absorber based on the highly absorbing wing scales of the \textit{Troides magellanus} Magellan birdwing butterfly. The characteristic geometry of the biological structure is rescaled and adapted to shift its optical response from the visible to the infrared spectral range. Numerical simulations are used to investigate the influence of the main geometrical parameters on the spectral absorptance and to identify an optimized configuration. The optimized structure achieves an absorption enhancement of 391\% compared with a flat steel reference. 
The improved optical response is attributed to the optimized high-aspect-ratio geometry, which enhances multiple scattering and light trapping, thereby increasing the effective light-matter interaction area. These results demonstrate the potential of bioinspired geometric scaling as an effective strategy for engineering efficient infrared absorbers and tailoring their spectral response for thermal radiation management and infrared photonic applications.

\end{abstract}

\section{Introduction}

Heat exchange through thermal radiation is an inherent feature of both natural and technological systems. Whenever a surface reaches a temperature above its surroundings, part of its thermal energy is inevitably emitted as electromagnetic radiation. This phenomenon occurs across a broad range of temperatures and is therefore found in numerous situations, from the solar-heated surface of the Earth to engines, industrial machinery, transportation systems, and household appliances. 

A significant fraction of this emission lies within the mid-infrared (mid-IR) spectral range, making thermal radiation an abundant energy source that remains largely unused. In practical applications, temperatures typically encountered in heat-generating systems can range approximately from $150$°C to $950$°C, resulting in significant radiative power being continuously emitted. Rather than being recovered, this energy is commonly released into the surrounding environment as part of the thermal losses associated with operation. Recovering this radiative energy and converting it into electricity could provide an additional way to improve the energy efficiency of existing systems, while potentially enabling part of their energy demand to be met by their own thermal emissions.

Efficient recovery of radiative energy requires effective control over the interaction between electromagnetic radiation and matter. Structured photonic materials provide a means to tailor this interaction by directing, reflecting, or confining incident radiation within specific regions of a device~\cite{Yablonovitch1982,Niggemann2008,Herman2012,Brongersma2014,Mayer2014,Deparis2015,Mayer2015,Rahman2015,Madanu2022,Madanu2023a,Madanu2023b,Wang2024}. This concept has been widely explored in energy-related applications, including photovoltaics, thermophotovoltaics, solar-thermal systems, thermoelectric generators (TEGs), artificial photosynthesis, and photocatalysis. The ability to engineer the optical response of a material therefore offers a powerful means of increasing the amount of radiation available for energy conversion.

Natural systems offer an additional source of inspiration for the development of efficient photonic devices. Many organisms have evolved highly effective ways of interacting with and controlling electromagnetic radiation, with these functions often playing an important role in their survival and adaptation~\cite{Mouchet2025,Ratoni2025}. Over millions of years of evolution, biological systems have developed sophisticated structural features that can enhance their interaction with electromagnetic radiation~\cite{mouchet2018structural,Mouchet2021}. These naturally occurring surfaces and integuments have consequently attracted interest as models for the design and fabrication of advanced, energy-efficient materials~\cite{Biro2011,Deparis2014,Zhou2017,Mouchet2021,Jayamaha2025,Delmote2026}. Bioinspired design therefore provides a promising approach for transferring functional principles observed in biological systems to engineered optical materials.

Biological surfaces found in fish, birds, and insects can act as efficient collectors and managers of thermal radiation through their specific structural organization~\cite{Biro2011,Zhou2017,Yoshida2002,Vasiljevic2021}. Beyond their ability to absorb thermal radiation, these structures can provide a combination of optical functionalities, including structural coloration produced by nanoscale interference effects~\cite{Mouchet2012b,Pasteels2016,kinoshita2008structural,Vukusic2001,Vukusic2003,mouchet2018IF,mouchet2018structural,Mouchet2021,Camus2026}, broadband antireflection~\cite{Yoshida2002,Stavenga2006,Deparis2014}, thermoregulatory behavior~\cite{Biro2003,Berthier2005,StuartFox2017,Liu2019,Xie2019,Mouchet2025}, efficient light trapping~\cite{Herman2011,Zhao2011,Vukusic2004,Han2015,Shanks2015,Tian2015,Davis2020}, and enhanced light extraction~\cite{Bay2013}. These different optical functions result from the interaction between electromagnetic radiation and complex micro- and nanoarchitectures. Such architectures are often formed from natural polymers, including chitin, keratin, collagen, and cellulose, and may also incorporate porous features.

Understanding how these structures achieve their optical functions can provide useful guidelines for the design of artificial photonic systems. Their properties do not result solely from the choice of material, but also from the precise organization of simple building blocks across several length scales. This combination of composition, geometry, and structural hierarchy offers opportunities to reproduce selected optical effects in synthetic materials while adapting them to specific technological requirements. For this reason, the study of biological photonic structures has become an active area of research and has increasingly contributed to the development of bioinspired optical technologies~\cite{Biro2011,Deparis2014,Zhou2017,Mouchet2021}.

\begin{figure}[htbp]
    \centering
    \includegraphics[width=1\linewidth]{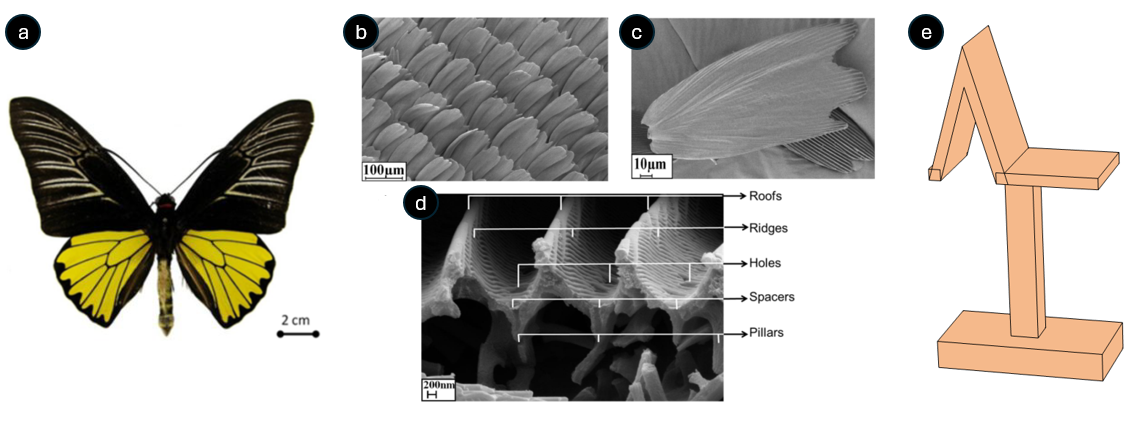}
    \caption{The male \textit{T. magellanus} butterfly, the Magellan birdwing (a), exhibits black forewings that absorb 98\% of the visible solar spectrum. This high absorption is associated with the presence of scales covering the wings. Scanning electron microscopy (SEM) images provide detailed views of the scale morphology at different magnifications. Image (b) shows the overall arrangement and overlapping of the scales on the wing surface. Image (c) provides a closer view of an individual scale, revealing its characteristic elongated morphology and surface organization. Image (d) shows a higher-magnification view of the scale structure, highlighting the nanoscale features within the scale. These hierarchical nanostructures are responsible for the optical properties of the butterfly wing and served as the biological inspiration for the absorber design. A 3D structure model comprising the morphological features giving rise to the highest absorptance spectrum was developed by Herman and coworkers~\citenum{Herman2011}(e). Figures (a-d) were reproduced from reference~\citenum{Herman2011}, with permission from SPIE.}    \label{fig:panel3}
\end{figure}

In this study, we focus on the \textit{Troides magellanus} birdwing butterfly, which is native to the Philippines and Orchid Island (Lanyu), Taiwan.
Beyond its impressive size, \textit{T. magellanus} exhibits remarkable differences in the optical properties of its forewings and hindwings (Figure~\ref{fig:panel3}). 
The forewings absorb approximately 98\% of incident visible light, due to their specific nanostructures and their composition, mainly involving chitin and melanin~\cite{Mouchet2025,Herman2011,Zhao2011,Davis2020}. This remarkable absorptance results from the combined action of reduced surface reflection, multiple scattering, light trapping, increased optical path length, and the enlarged effective interaction area provided by the hierarchical wing architecture, together with absorption by melanin and chitin~\cite{Mouchet2025}.
The chitin present in the forewings also gives rise to two absorption peaks in the infrared region, at approximately $3$~µm and $6$~µm. 
These peaks are associated with vibrational modes of carbonyl ($\textrm{C}=\textrm{O}$) groups and contribute to radiative cooling~\cite{Mouchet2025, Herman2011}.
In contrast, the hindwings combine two distinct optical phenomena. First, pigment-induced yellow-green fluorescence gives the hindwings their bright yellow coloration~\cite{VanHooijdonk2012b} and, second, a blue-green iridescence results from light diffraction under grazing illumination~\cite{Vigneron2008,Lee2009,VanHooijdonk2012}. 

In this context, we took inspiration from the nanostructures occurring in the forewings of male \textit{T. magellanus}. 
Herman \textit{et al.} examined the butterfly wings using scanning electron microscopy~\cite{Herman2011}. This work revealed the structural features of the wing morphology, including a factory-roof-shaped upper lamina and pillars separating the laminae, together with their dimensions.
Herman \textit{et al.} demonstrated that optical absorptance in the integuments of this butterfly does not depend solely on the chemical composition of the wings, particularly their chitin and melanin content, but also on their structural organization at the nanoscale~\cite{Herman2011}.
They managed to discriminate which morphological features give rise to the highest optical absorptance in the visible-light range and established a related structure model (Figure~\ref{fig:panel3}e). However, this simplified model does not fully capture the morphology observed in the natural scales.
In particular, the natural structures contain nanoscale perforations that are absent from the proposed model (Figure~\ref{fig:panel3}d).
Although the presence of nanoscale perforations slightly reduces absorption, it decreases wing mass and therefore represents a trade-off between optical performance and weight.

In this article, we improve mid-IR light harvesting by developing a bioinspired steel absorber modeled on the exceptionally dark and strongly absorbing wing surfaces of \textit{T. magellanus} birdwing butterflies. Exploiting the scale invariance of Maxwell's equations~\cite{Delmote2026}, biological architectures known for their remarkable visible-light absorption were geometrically rescaled and subsequently optimized to achieve efficient absorption in the mid-IR spectral range in a steel microstructured absorber.

\section{Materials and Methods}
Owing to the scale invariance of Maxwell’s equations, the dimensions of the nanostructures present on the wings of \textit{T. magellanus}, which are responsible for efficient visible-light absorption, can be adapted to operate in the mid-IR spectral range~\cite{Delmote2026}. 
This can be achieved by applying a multiplicative scaling factor to the characteristic dimensions of the biological structure. 

As the butterfly wing absorbs sunlight primarily within the visible spectral range (approximately 0.4-0.7~µm), its nanostructures can be approximated as operating under illumination from a blackbody source at a temperature $T$ of $5500$~K~\cite{Herman2011}, which is similar to a $\textrm{D}_{55}$ illuminant. In contrast, the bioinspired structure was designed to absorb thermal radiation emitted by a blackbody at $550$°C, or $823.15$~K. Applying such a scaling enlarges the characteristic dimensions of the structure from the submicrometre range to the micrometre scale, thereby facilitating its fabrication and patterning. 

Using Wien's displacement law
\begin{equation}
    \lambda_{\max} = \frac{\sigma_{\mathrm{W}}}{T},
\end{equation}
with $\sigma_{\mathrm{W}} = 2.89777291 \times 10^{-3}$~m~K, the scaling factor required to shift the optical response of the butterfly-inspired structure toward the mid-IR range is given by
\begin{equation}
    \gamma = \frac{\lambda_{823.15\,\mathrm{K}}}{\lambda_{5500\,\mathrm{K}}} \cdot \frac{n_{\mathrm{eff}}^{5500\,\mathrm{K}}}{n_{\mathrm{eff}}^{823.15\,\mathrm{K}}} \simeq 6.68.
\end{equation}
Here, $n_{\mathrm{eff}}^{5500~\mathrm{K}}$ and $n_{\mathrm{eff}}^{823.15~\mathrm{K}}$ denote the effective refractive indices of the chitin-based biological structure and the steel-based bioinspired structure, respectively.

Although this scaling procedure shifts the spectral response of the bioinspired architecture toward the targeted mid-IR range, it does not necessarily provide optimal absorption performance. The geometrical parameters were therefore further adjusted to maximize absorptance. To identify the most favorable configuration, a systematic parametric study was conducted in which absorptance spectra were calculated for different combinations of geometrical parameters using a brute-force optimization approach.
To compare the simulated spectra, we used an absorption coefficient $\alpha(T)$ corresponding to a spectrally weighted absorptance, as figure of merit:
\begin{equation}
\alpha(T) =
\frac{
\displaystyle \int_{\lambda_1}^{\lambda_2}
\varepsilon_{\mathrm{BB}}(\lambda,T) \mathcal{A}(\lambda)\,\mathrm{d}\lambda
}{
\displaystyle \int_{\lambda_1}^{\lambda_2}
\varepsilon_{\mathrm{BB}}(\lambda,T)\,\mathrm{d}\lambda
},
\label{eq:abs}
\end{equation}
where $\varepsilon_{\mathrm{BB}}(\lambda,T)$ is the spectral emissive power of a blackbody at temperature $T$ and $\mathcal{A}(\lambda)$ is the absorptance spectrum of the investigated structure model. The blackbody temperature was set to $T = 823.15$~K, corresponding to the temperature of the heat source.
The integrals were evaluated over the wavelength range from $0.67$~µm to $18.2$~µm. The blackbody emissivity spectrum was calculated using Planck’s law.
In the following analysis, the improvement in absorption relative to a reference model consisting of an arbitrarily 3.58~$\mu$m-thick flat steel slab was quantified using the enhancement factor $A_{\mathrm{enh}}(T)$ defined as:

\begin{equation}
A_{\mathrm{enh}}(T) = \frac{\alpha_{\mathrm{m}}(T)-\alpha_{\mathrm{r}}(T)}{\alpha_{\mathrm{r}}(T)},
\label{eq:absorption_enhancement}
\end{equation}
where $\alpha_{\mathrm{m}}(T)$ is the absorption coefficient of the investigated model and $\alpha_{\mathrm{r}}(T)$ is the absorption coefficient of the reference model. 

The complete optimization procedure was carried out in silico using finite-element electromagnetic simulations. All simulations were performed using COMSOL Multiphysics 6.2, which solves Maxwell’s equations using the finite element method (FEM), on a workstation equipped with an Intel® Core™ i9-14900K processor with 24 cores. This numerical approach is particularly well suited to modelling diffraction, propagation, absorption, and scattering phenomena in structures with complex geometries. 

To determine an appropriate mesh discretization, a convergence study was performed. The final mesh size was set to $\frac{\lambda}{15.n_{\mathrm{steel}}(\lambda)}$. This mesh was then used throughout all simulations to ensure consistency and enable reliable comparisons between the different geometrical configurations.
The computation time for each simulation ranged from approximately 10 to 30 minutes, depending on the geometrical configuration.  

\section{Results and Discussion}

Steel was selected as the constituent material of the bioinspired absorber because it can be readily microstructured using techniques such as laser metal machining (LMM)~\cite{MELENTIEV202580} and femtosecond laser ablation~\cite{bruneel:tel-00675144}. 
In addition, steel exhibits a relatively high thermal conductivity $\kappa$, which is advantageous for the targeted energy-harvesting application, as it facilitates the efficient transfer of the absorbed thermal energy~\cite{haynes2014crc}.

\begin{figure}[H]
    \centering
    \includegraphics[width=0.75\linewidth]{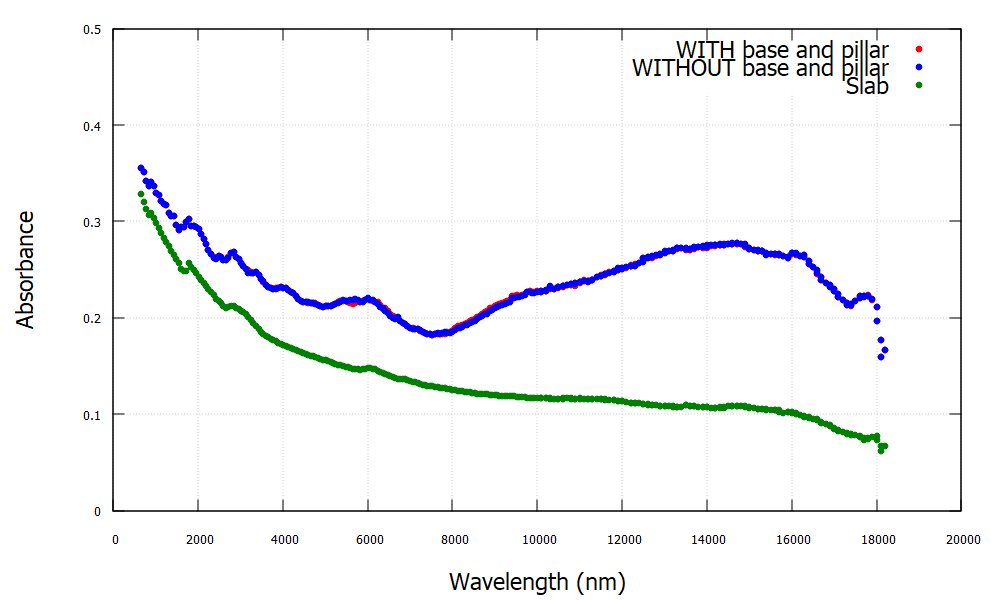}
    \caption{The absorptance spectra of the structure including the base and pillar (red curve) and of the simplified structure without the base and pillar (blue curve) are superimposed, indicating that the electromagnetic waves do not propagate through the factory-roof-shaped element of the structure. Therefore, the base and pillar have a negligible contribution to the overall absorption and can be removed from the model to simplify the simulations, and the microfabrication of the structure. The scaled structures exhibit a higher spectral absorptance than an arbitrarily 3.58~$\mu$m-thick flat steel slab (green curve).}
    \label{fig:comp}
\end{figure}

\begin{figure}[htbp]
    \centering
    \begin{tikzpicture}
    \node (img1) at (0,0) {
        \includegraphics[width=0.4\textwidth]{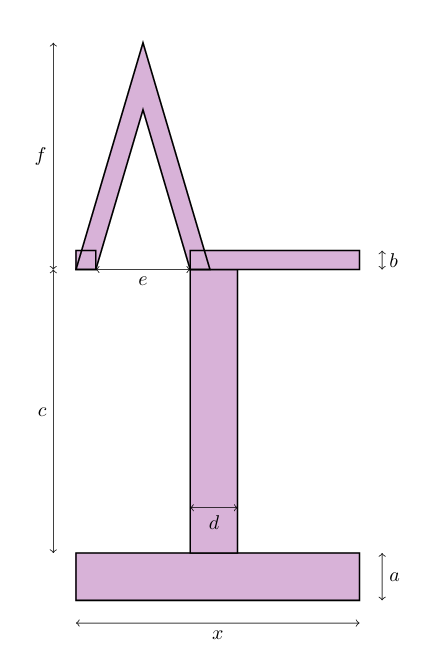}
    };
    \node[anchor=north west] at (img1.north west) {\textbf{(a)}};
    \node (img2) at (8,0) {
        \includegraphics[width=0.45\textwidth]{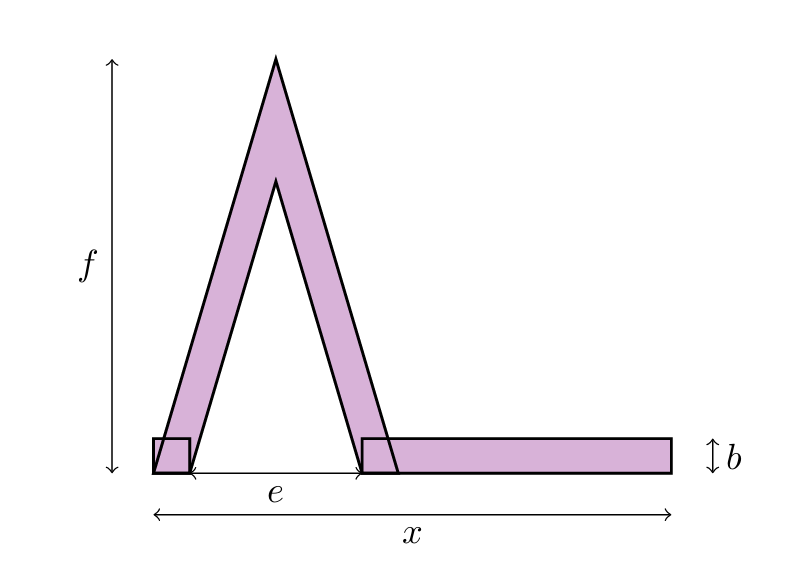}
    };
    \node[anchor=north west, yshift=2.39cm] at (img2.north west) {\textbf{(b)}};
        \draw[->, very thick] 
            (2.5,0) -- (4.8,0);
    \end{tikzpicture}
    \caption{Because the metallic structure prevents electromagnetic wave propagation beyond the upper membrane, the base and pillar can be removed without altering the calculated absorptance. The resulting simplification significantly reduces the computational complexity of the model. The initial up-scaled structure and the simplified bioinspired geometry are shown in (a) and (b), respectively. The former consists of a lower membrane of thickness $a$, an upper membrane of thickness $b$ with a triangular feature defined by a base $e$ and a height $f$. The separation between the two membranes is denoted by $c$, the supporting pillar has a width $d$, and the lattice period is $x$.}
    \label{fig:AVECSANS}
\end{figure}

The scaled non-optimized structure model already exhibits higher absorptance than a flat steel surface (Figure~\ref{fig:comp}), demonstrating the beneficial effect of the biomimetic geometry prior to optimization. Owing to the metallic nature of the structure, electromagnetic fields remain confined to the vicinity of the upper membrane and do not propagate beyond the factory-roof-shaped feature (Figure~\ref{fig:comp}). Consequently, the pillar and the base have a negligible influence on the absorptance. This observation makes it possible to replace the full 3D geometry by a simplified 2D representation (Figure~\ref{fig:AVECSANS}) without loss of accuracy in the optical simulations. Besides reducing the computational burden, this simplification also decreases the geometrical complexity of the structure from a fabrication perspective.

Following the geometrical simplification of the model, only lattice period $x$ of the structure, the thickness $b$ of the lamina, the height $f$ of the triangular feature, and the base $e$ of the triangular feature were left for the parametric optimization (Table~\ref{tab:parameters_description}).

\begin{table}[H]
    \centering
    \begin{tabular}{|c|l|c|c|}
        \hline
        \textbf{Parameter} & \textbf{Description} & \textbf{Initial value} (µm) & \textbf{Optimized value} (µm) \\
        \hline
        $x$ & Lattice period of the structure & $10.0$ & $4.80$ \\
        \hline
        $b$ & Thickness of the lamina & $0.67$ & $0.24$ \\
        \hline
        $f$ & Height of the triangular feature & $8.0$ & $25.0$ \\
        \hline
        $e$ & Base of the triangular feature & $3.33$ & $3.10$ \\
        \hline
    \end{tabular}
    \caption{Geometrical parameters $x$, $b$, $f$, and $e$ used in the optimization process, together with their initial and optimized values.}
    \label{tab:parameters_description}
\end{table}

\begin{figure}[htbp] 
    \centering 
    \begin{subfigure}[b]{0.48\textwidth} 
        \centering 
        \begin{tikzpicture} 
            \node[inner sep=0] (img1) {\includegraphics[width=\textwidth]{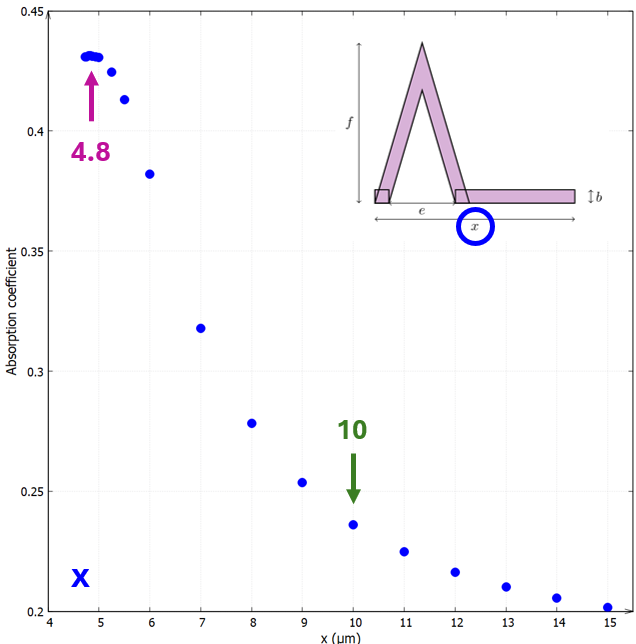}};
            \node at ([xshift=-0.1cm,yshift=0.1cm]img1.north west) {\textbf{(a)}};
        \end{tikzpicture} 
    \end{subfigure} 
    \hfill 
    \begin{subfigure}[b]{0.48\textwidth} 
        \centering 
        \begin{tikzpicture} 
            \node[inner sep=0] (img2) {\includegraphics[width=\textwidth]{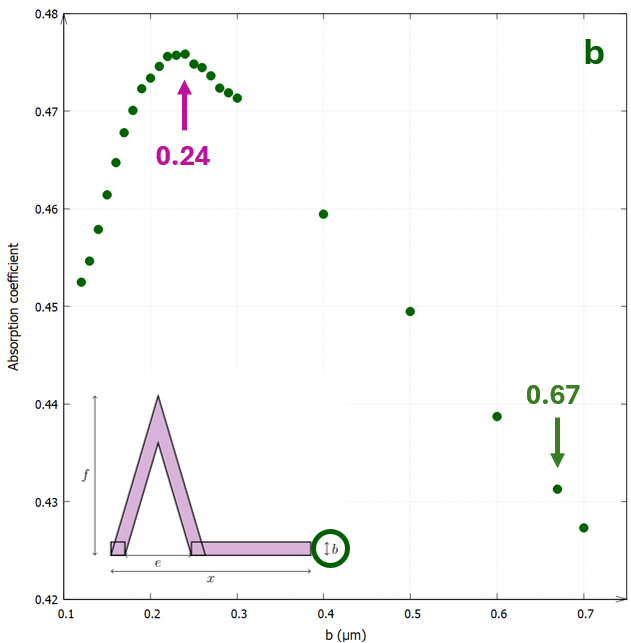}}; 
            \node at ([xshift=0cm,yshift=0.05cm]img2.north west) {\textbf{(b)}}; 
        \end{tikzpicture} 
    \end{subfigure} 
    \vspace{0.4cm}
    \begin{subfigure}[b]{0.48\textwidth} 
        \centering 
        \begin{tikzpicture} 
            \node[inner sep=0] (img3) {\includegraphics[width=\textwidth]{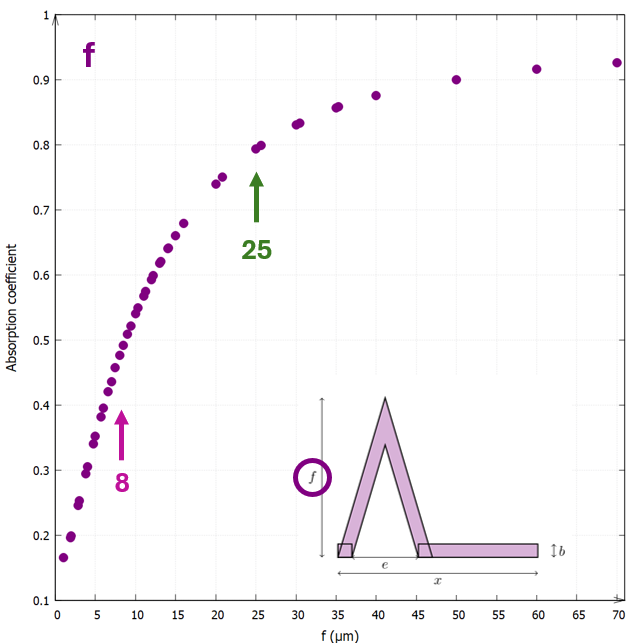}}; 
            \node at ([xshift=0.1cm,yshift=0.11cm]img3.north west) {\textbf{(c)}}; 
        \end{tikzpicture}
    \end{subfigure} 
    \hfill 
    \begin{subfigure}[b]{0.48\textwidth} 
        \centering 
        \begin{tikzpicture}
            \node[inner sep=0] (img4) {\includegraphics[width=\textwidth]{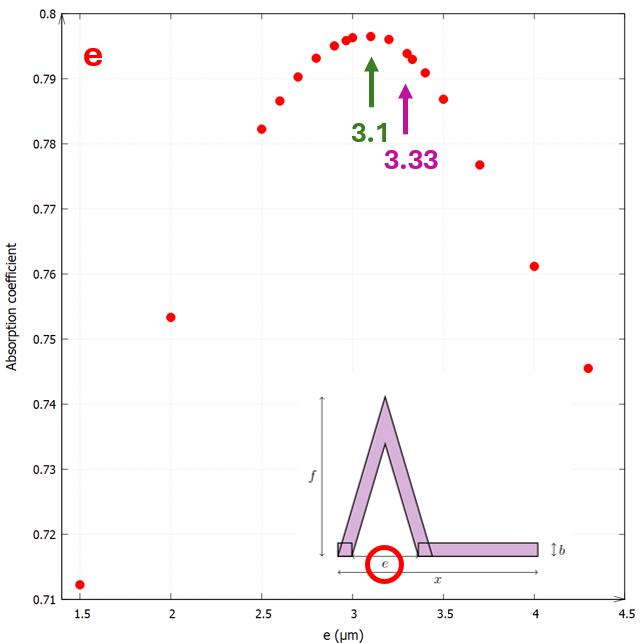}}; 
            \node at ([xshift=0.1cm,yshift=0.14cm]img4.north west) {\textbf{(d)}}; 
        \end{tikzpicture} 
    \end{subfigure} 
    
    \caption{The optimization of the geometrical parameters of the bioinspired absorber led to optimal values of $x=4.80$~µm, $b=0.24$~µm, $f=25$~µm and $e=3.10$~µm, respectively. (a) Variation of the absorption coefficient $\alpha(T=823.15\,\mathrm{K})$ as a function of the period $x$, yielding a maximum at $x=4.80$~\textmu m. (b) Variation of $\alpha(T=823.15\,\mathrm{K})$ as a function of the upper lamina thickness $b$ after optimization of $x$, yielding a maximum at $b=0.24$~\textmu m. (c) Variation of $\alpha(T=823.15\,\mathrm{K})$ as a function of the triangular feature height $f$ after optimization of $x$ and $b$, showing an increase in absorptance. (d) Variation of $\alpha(T=823.15\,\mathrm{K})$ as a function of the triangular feature base width $e$ after optimization of $x$, $b$, and $f$, yielding a maximum at $e=3.10$~\textmu m. } 
    
    \label{fig:optimizationcomparison} 
    \end{figure}

The optimization process was initiated by investigating the influence of the period $x$ as this parameter governs the overall dimensions of the structure and directly affects the remaining geometrical features. The period was varied between $4.73$~µm and $15.0$~µm (Figure~\ref{fig:optimizationcomparison}a). The lower bound was determined by the requirement that the triangular upper feature remain fully defined within the unit cell, whereas the upper bound was selected based on the observed monotonic decrease in absorptance with increasing period. The results revealed that the absorptance reaches a peak at a period below approximately $5$~µm, with a maximum obtained for $x = 4.80$~µm (Figure~\ref{fig:optimizationcomparison}a). Consequently, this value was retained for all subsequent simulations. The resulting structure exhibited an absorption enhancement $A_{\mathrm{enh}} = 166.1\%$ relative to the unstructured steel surface, highlighting the strong sensitivity of the absorber performance to the lattice period.

The lamina thickness, $b$, was found to strongly influence the absorptance of the structure (Figure~\ref{fig:optimizationcomparison}b). In addition to its potential effect on the optical performance, this parameter is of practical importance for the fabrication of the structure. The membrane thickness was varied between $0.12$~µm and $0.70$~µm. The lower bound was determined by the minimum thickness for which the mesh remained sufficiently refined to ensure accurate simulations, whereas the upper bound was selected because the absorptance was found to decrease monotonically (Figure~\ref{fig:optimizationcomparison}b). A maximum absorptance was obtained for a lamina thickness of $b=0.24$~µm (Figure~\ref{fig:optimizationcomparison}b), indicating the existence of an optimal membrane thickness for light harvesting. This value was therefore retained for the subsequent optimization steps. The optimized configuration yielded $A_{\mathrm{enh}}=193.59\%$, demonstrating a substantial improvement in absorptance with respect to the reference configuration.

The effect of the structure height, $f$, on the absorptance was subsequently examined. The parameter was varied from $1.0$~$\mu$m to $70.0$~$\mu$m. The lower bound was selected because, below this value, the triangular roof effectively behaves as a planar surface and no longer exhibits its characteristic geometry. The upper bound was chosen because the absorptance approaches a plateau beyond this range (Figure~\ref{fig:optimizationcomparison}c).
Although the highest absorptance values were obtained for larger heights, a value of $f = 25.0$~$\mu$m was retained for the final design. This height lies close to the absorptance plateau while remaining substantially more practical from a fabrication perspective than taller structures. Consequently, it represents a practical compromise between absorptance enhancement and fabrication constraints. 
Using this optimized value, absorptance enhancement of $A_{\mathrm{enh}}=382\%$ relative to the reference steel flat slab was obtained. These results highlight the significant contribution of the structure height to the overall absorption performance.

The parameter $e$ exhibited a non-monotonic influence on the absorptance (Figure~\ref{fig:optimizationcomparison}d), leading to the emergence of a distinct optimum at $e = 3.10$~µm. Absorptance decreased progressively when the parameter was varied away from this value in either direction, indicating that efficient light trapping requires a precise geometrical balance. The optimization process therefore retained $e = 3.10$~µm for the final structure. 
The resulting design achieved absorptance enhancement of $A_{\mathrm{enh}}=391.39\%$ relative to the steel flat reference. This value corresponds to the highest performance obtained throughout this brute-force optimization study.

\begin{figure}[htbp]
    \centering
    \begin{tikzpicture}
        \node[inner sep=0] (img1) at (0,0) {
            \includegraphics[width=0.45\textwidth]{FIG_SANS.png}
        };
        \node at ([xshift=0.1cm,yshift=1.82cm]img1.north west) {\textbf{(a)}};

        \node[inner sep=0] (img2) at (8,0) {
            \includegraphics[width=0.2\textwidth]{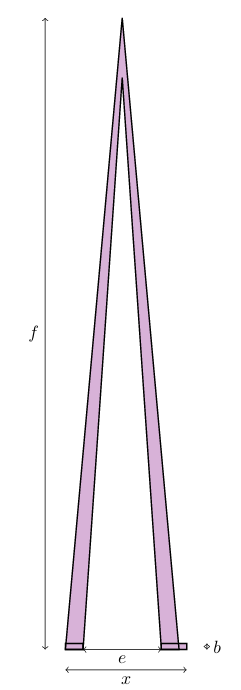}
        };
        \node at ([xshift=-0.3cm,yshift=-0.2cm]img2.north west) {\textbf{(b)}};
        \draw[->, very thick] 
            (3,0) -- (6,0);
    \end{tikzpicture}

    \caption{Optimization drives the bioinspired absorber toward a high-aspect-ratio architecture that improves optical absorption. Comparison between the initial geometry obtained through electromagnetic scaling (a) and the final optimized structure (b). The optimized design is taller and narrower, favouring multiple scattering and enhanced interaction between the incident radiation and the absorbing material.}
    \label{fig:initialoptimized}
\end{figure}

The optimization procedure resulted in a slender, high-aspect-ratio architecture characterized by a reduced lateral footprint and an increased height (Figure~\ref{fig:initialoptimized}). Relative to the initially scaled geometry, the optimized structure provides a larger effective light-matter interaction area and promotes multiple scattering of the incident radiation within the absorber. These effects increase the optical path length and enhance the interaction between the electromagnetic field and the absorbing material, leading to a substantial improvement in absorptance.

\begin{figure}[H]
    \centering
    \includegraphics[width=0.75\linewidth]{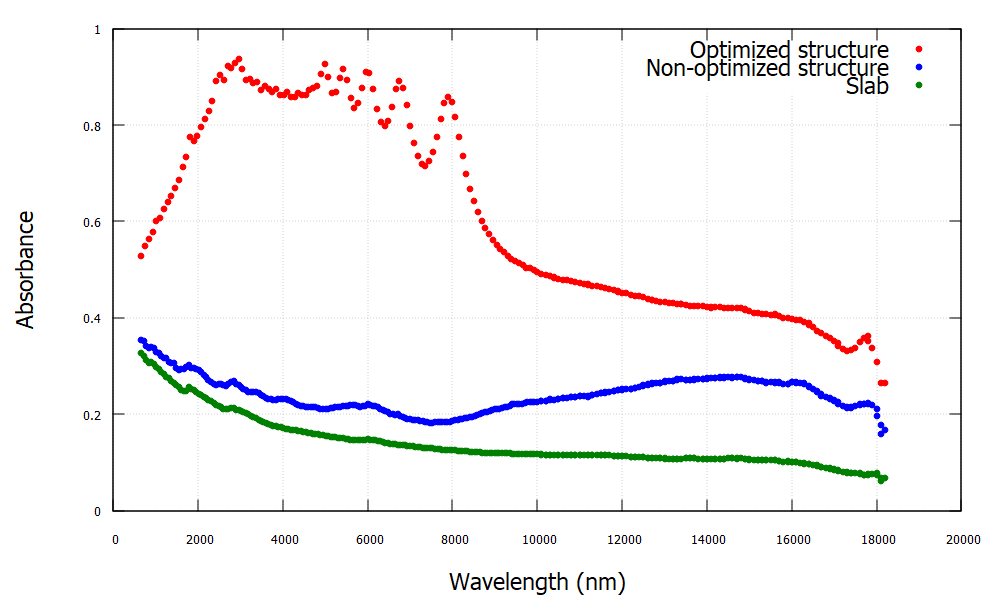}
    \caption{The optimized bioinspired absorber exhibits significantly improved broadband absorptance relative to both the initial non-optimized structure and the flat steel slab. Absorptance spectra of the steel slab, the initial bioinspired geometry, and the optimized structure in the 0.67-18.20~µm wavelength range. The optimization results in a pronounced increase in absorptance, particularly between 2 and 10~µm.}
    \label{fig:final}
\end{figure}

A substantial enhancement in absorptance is achieved through geometrical optimization (Figure~\ref{fig:final}). Compared with both the flat steel reference and the initial non-optimized scaled geometry, the optimized structure exhibits markedly improved absorption, reaching values approximately nine and four times greater, respectively. The enhancement is particularly pronounced in the short-wavelength portion of the spectrum, below approximately 10~µm, leading to a significant increase in broadband absorptance.

\begin{figure}[H]
    \centering
    \includegraphics[width=0.75\linewidth]{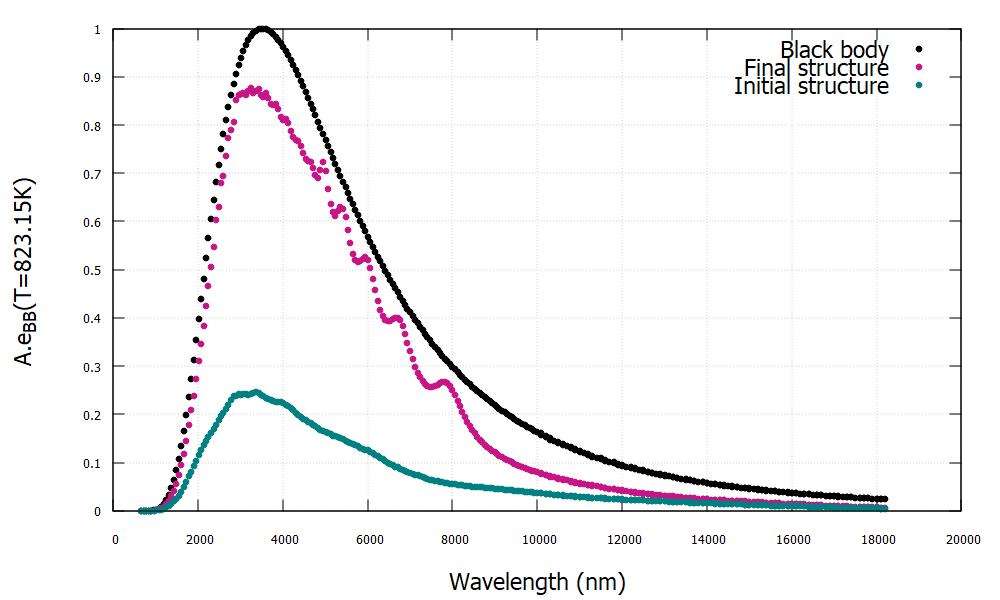}
    \caption{The optimized structure exhibits nearly four times greater absorptance than the initial structure near the maximum of the blackbody emission spectrum. Blackbody spectral emissive power at $T=823.15$~K together with the corresponding $\mathcal{A}\times\epsilon_{\mathrm{BB}}(\lambda,T=823.15\mathrm{~K})$ for the non-optimized and optimized bioinspired structures as a function of wavelength (nm).}
    \label{fig:BB}
\end{figure}

The importance of the 2-10~µm region is further highlighted by the blackbody emission spectrum of the heat source (Figure~\ref{fig:BB}), which exhibits a maximum at 3.52~µm. Consequently, absorptance enhancements achieved near this wavelength contribute disproportionately to the integrated absorption coefficient. The strong increase in absorptance exhibited by the optimized structure in this spectral range therefore plays a decisive role in its superior overall performance (Figure~\ref{fig:BB}).

The optimization of the structure favours a reduction of the structure period together with an increase in the height of the triangular feature, ultimately leading to a slender high-aspect-ratio architecture. Such geometries are known to promote light trapping through multiple scattering events and reduced reflectance, while simultaneously increasing the effective light-matter interaction area~\cite{Mouchet2025}. These mechanisms extend the optical path length within the absorber and increase the probability of absorption, thereby accounting for the superior performance of the optimized bioinspired structure.

\section{Conclusion}
The present study demonstrates that the optical response of the bioinspired absorber can be significantly enhanced through geometrical scaling and optimization. Owing to the metallic nature of the structure, electromagnetic waves do not propagate beyond the upper membrane, rendering the pillar and the underlying base largely irrelevant to the absorption process. This observation justifies the use of a simplified numerical model and enables a substantial reduction in computational cost without compromising accuracy.
Steel emerged as a particularly attractive candidate for infrared absorbers because of its strong absorptance, high thermal conductivity, and compatibility with established microfabrication techniques. The optimization procedure further revealed that high-aspect-ratio architectures considerably improve light trapping within the structure.
The optimized bioinspired design achieved an absorptance enhancement approaching 400\% relative to a flat steel reference. This improvement is attributed to the combined effects of multiple scattering, improved index matching, and an increase in the effective light-matter interaction area within the structure. These results demonstrate that geometrically scaled biological architectures can provide valuable design principles for the development of high-performance infrared photonic devices and energy-harvesting technologies.

\bibliographystyle{IEEEtran}
\bibliography{references}
\end{document}